\documentclass[twoside]{article}
\usepackage{pgfpages}
\usepackage{jp}
\pgfpagesuselayout{2 on 1 shifted}[border shrink=-1.5in,vertical shift=-.35in,horizontal shift=.175in,in shift=.25in]
\usepackage{amssymb}
\usepackage{amsmath}
\usepackage{amsfonts}

\usepackage{url}
\usepackage{color}
\usepackage{graphicx}

\usepackage[T1]{fontenc}
\usepackage{tikz}
\usetikzlibrary{arrows,shapes,calc}
\usetikzlibrary{shadows}
\DeclareMathAlphabet{\mathitbf}{OML}{cmm}{b}{it}

\title{Mechanizing \emph{Principia Logico-Metaphysica} in Functional
  Type Theory\thanks{Copyright \copyright\ 2019, by the authors. This
    paper is forthcoming at the \emph{Review of Symbolic Logic}.}}

\author{Daniel Kirchner\\Fachbereich Mathematik und Informatik \\Freie
  Universit\"at Berlin \\ {\tt daniel@ekpyron.org} \and Christoph
  Benzm\"uller \\ Fachbereich Mathematik und Informatik \\ Freie
  Universit\"at Berlin \\ {\tt c.benzmueller@fu-berlin.de} \\ \&
  Computer Science and Communications \\ University of Luxembourg
  \\ {\tt christoph.benzmueller@uni.lu} \\ \and Edward
  N. Zalta\\ Center for the Study of Language and
  Information\\ Stanford University \\ {\tt zalta@stanford.edu} }

\date{}

\begin{document}

\thispagestyle{empty}

\maketitle

\begin{abstract}
\emph{Principia Logico-Metaphysica} contains a foundational logical
theory for metaphysics, mathematics, and the sciences.  It
includes a canonical development of Abstract Object Theory [AOT], a
metaphysical theory (inspired by ideas of Ernst Mally, formalized by
Zalta) that distinguishes between ordinary and abstract objects.

This article reports on recent work in which AOT has been successfully
represented and partly automated in the proof assistant system
Isabelle/HOL. Initial experiments within this framework reveal a
crucial but overlooked fact: a deeply-rooted and known paradox is
reintroduced in AOT when the logic of complex terms is simply adjoined
to AOT's specially formulated comprehension principle for
relations. This result constitutes a new and important paradox, given
how much expressive and analytic power is contributed by having the
two kinds of complex terms in the system. Its discovery is the
highlight of our joint project and provides strong evidence for a new
kind of scientific practice in philosophy, namely, \emph{computational
  metaphysics}.

Our results were made technically possible by a suitable adaptation of
Benzm\"ul\-ler's metalogical approach to universal reasoning by
semantically embedding theories in classical higher-order logic. This
approach enables one to reuse state-of-the-art higher-order proof
assistants, such as Isabelle/HOL, for mechanizing and experimentally
exploring challenging logics and theories such as AOT. Our results
also provide a fresh perspective on the question of whether relational
type theory or functional type theory better serves as a foundation
for logic and metaphysics.

\end{abstract}

\section{Abstract Summary}

\emph{Principia Logico-Metaphysica} (PLM) [16] is an online
research monograph that contains a canonical presentation of Abstract
Object Theory (AOT) [17], [18],
along with motivation for, and commentary on, the theory.  AOT is a
foundational logical theory for metaphysics, mathematics and the
sciences.  It distinguishes between abstract and ordinary objects, by
regimenting a distinction found in the work of the philosopher Ernst
Mally [7] (though the distinction has appeared in other philosophical
works).

AOT is outlined in \S \ref{sec:aot}. It systematizes two fundamental
kinds of predication: classical exemplification for ordinary and
abstract objects, and \emph{encoding} for abstract objects.  The
latter is a new kind of predication that provides AOT with expressive
power beyond that of quantified second-order modal logic, and this
enables one to formalize various metaphysical theories about different
abstract objects, including the objects presupposed by mathematics and
the sciences. More generally, the system offers a universal logical
theory that may have a greater capability of accurately representing
the contents of human thought than other foundational systems.

Independently, the use of \emph{shallow semantical embeddings} (SSEs)
of complex logical systems {in classical higher-order logic} (HOL) has
consistently shown potential as a metalogical approach towards
{universal logical reasoning} [1].  The SSE approach aims to unify
logical reasoning by using HOL as a universal metalogic. Only the
distinctive primitives of a target logic are defined in the metalogic
in terms of their semantic interpretations, whereas the rest of the
target system is captured by the existing infrastructure of HOL. This
is why it is a \emph{shallow} semantical embedding. For example,
quantified modal logic can be encoded by representing propositions as
sets of possible worlds and by representing the connectives,
quantifiers, and modal operators as operations on those sets. In this
way, the world-dependency of Kripke-style semantics can easily be
modeled in HOL. Utilizing the variety of options for handling and
hiding such definitions that are offered in modern proof assistants
such as Isabelle/HOL [12], a human-friendly mechanization of even the
most challenging target logics, including AOT, can thus be obtained.

AOT and the SSE approach are rather orthogonal. They have very
different motivations and come with fundamentally different
foundational assumptions. AOT uses a \emph{hyperintensional
  second-order modal logic}, grounded on a \emph{relational type
  theory}, as its foundation. It is in the tradition of Whitehead and
Russell's \emph{Principia Mathematica} [14], [10] which takes the
notion of \emph{relation} as primitive and defines the notion of
\emph{function} in terms of relations. Relations, on this approach,
are assumed to be hyperintensional unless one explicitly asserts, of a
given $n$-place relation $R$, that it is extensional by adding an
axiom that guarantees that any $n$-place relation necessarily
equivalent to $R$ is identical to it, i.e., that $\forall
S(\Box\forall x_1\ldots \forall x_n(Sx_1\ldots x_n \equiv Rx_1\ldots
x_n) \to S\! =\! R)$. The metalogic HOL in the SSE approach, by
contrast, is fully extensional, and is defined on top of a functional
type theory in the tradition of the work of Frege [6] and Church
[4]. It takes the notion of (fully extensional) \emph{function} as
primitive and defines the notion of \emph{relation} in terms of
functions.\footnote{In principle, functional type theory can be
  developed intensionally~[11], but Isabelle/HOL and other automated
  proof assistants are based on extensional functional type theory for
  a reason, namely, to make the system computationally tractable.}
These fundamentally different and, to some extent, antagonistic roots
impose different requirements on the corresponding frameworks, in
particular, with regard to the comprehension principles that assert
the existence of relations and functions. Devising a mapping between
the two formalisms is, unsurprisingly, a non-trivial, practical
challenge [13].

The work reported here tackles this challenge. Further details can be
found in Kirchner's M.A.\ thesis~[8], where the SSE approach
is utilized to mechanize and analyze AOT in HOL. Kirchner constructed
a shallow semantical embedding of the second-order modal fragment of
AOT in HOL, and this embedding was subsequently represented in the
proof assistant system Isabelle/HOL
(see~\S\ref{sec:aotinIsabelle}). The proof assistant system enables
one to conduct experiments in the spirit of a \emph{computational
  metaphysics}, with results that have helped to advance the ideas of
AOT.

The inspiration for Kirchner's embedding comes from the model for AOT
proposed by Peter Aczel.\footnote{An earlier model for AOT was
  proposed by Dana Scott. His model is equivalent to a special case of
  an Aczel model with only one \emph{special urelement}.  See below
  for a discussion of the Aczel model.}  Kirchner adapted techniques
used in Benzm\"uller's initial attempts to embed AOT in Isabelle/HOL.
An important goal of the research was to avoid \emph{artifactual
  theorems}, i.e., theorems that (a) are derivable on the basis of
special facts about the Aczel model that was used to embed AOT in
Isabelle/HOL, but (b) aren't theorems of AOT.\footnote{We have not yet
  investigated the question of whether the embedding of AOT in HOL is
  complete in the sense that if the representation of $\phi$ is
  provable in HOL, then $\phi$ is provable in AOT. However, this will
  be a topic of future research.} An example of an artifactual theorem
can be seen in Figure~\ref{fig:artifactual} and, since AOT is, in
part, a body of theorems, care has been taken not to derive
artifactual theorems about the Aczel model that are not theorems of
AOT itself.

This explains why the embedding of AOT in Isabelle/HOL involves
several layers of abstraction. In the Aczel model of AOT that serves
as a starting point, abstract objects are modeled as sets of
properties, where properties are themselves modeled as sets of
urelements. Once the axioms of AOT are derived from the shallow
semantical embedding of AOT in HOL, a controlled, and suitably
restricted, logic layer is defined.  By reconstructing the inference
principles of AOT in the system that derives the axioms of AOT, only
the theorems of AOT become derivable. If we utilize Isabelle/HOL's
sophisticated support tools for interactive and automated proof
development at this highest level of the embedding, it becomes
straightforward to map the pen and paper proofs of PLM into
corresponding, intuitive, and user-friendly proofs in Isabelle/HOL. In
nearly all cases this mapping is roughly one-to-one, and in several
cases the computer proofs are even shorter. In other words, the
\emph{de Bruijn factor} [15] of this work is close to
1. In addition, the layered construction of the embedding has yielded
a detailed, experimental analysis in Isabelle/HOL of the underlying
Aczel model and the semantic properties of AOT.

As an unexpected, but key result of this experimental study, it was
discovered that if a classical logic for complex terms such as
$\lambda$-expressions and definite descriptions is adjoined, without
taking any special precautions, to AOT's specially-formulated
comprehension principle for relations, a known paradox that had been
successfully put to rest is reintroduced (see \S
\ref{sec:paradox}). Since the complex terms add significant expressive
and analytic power to AOT, and play a role in many of its more
interesting theorems and applications, the re-emergence of the known
paradox has become a \emph{new} paradox that has to be addressed.  In
the ongoing attempts to find an elegant formulation of AOT that avoids
the new paradox, the computational representation in Isabelle/HOL now
provides a testing infrastructure and serves as an invaluable aid for
analyzing various conjectures and hypothetical solutions to the
problem. This illustrates the very idea of \emph{computational
  metaphysics}: humans and machines team up and split the tedious work
in proportion to their cognitive and computational strengths and
competencies. And, as intended, the results we achieved reconfirm the
practical relevance of the SSE approach to universal logical
reasoning.

% and the embedding can continue to serve as a sophisticated tool to
% explore possible options to adjust the theory to avoid the
% discovered paradox.

Though the details of the embedding of AOT in Isabelle/HOL are
developed in Kirchner [8], we discuss the core aspects of
this work in the remainder of this article.

\section{The Theory of Abstract Objects} \label{sec:aot}

The language of second-order AOT uses individual variables
$x,y,\ldots$ and $n$-place relation variables $F^n,G^n,\ldots$ ($n\geq
0$).  The sentences of the language are built up from two kinds of
atomic formulas: classical \emph{exemplification} formulas of the form
$F^1x$ (or more generally, $F^nx_1\ldots x_n$) and \emph{encoding}
formulas of the form $xF^1$.\footnote{It is important to emphasize
  that the two kinds of atomic formulas are \emph{typed} in the sense
  that, in both kinds of formulas, no individual term can stand in
  relation position and no relation term can stand in individual
  position.  In the embedding of this language in functional type
  theory both kinds of formulas are given a special functional
  semantics that is distinct from the fundamental notion of function
  application in HOL. Details of this semantics can be found in
  \S\ref{sec:aotinIsabelle}.} A distinguished predicate is then
used to define a distinction between objects that exemplify
\emph{being abstract} ($A!x$) and objects that exemplify \emph{being
  ordinary} ($O!x$). Whereas ordinary objects are characterized only
by the properties they exemplify, abstract objects may be
characterized by both the properties they exemplify and the properties
they encode. But only the latter play a role in their identity
conditions: $A!x \:\&\: A!y \to \allowbreak (x\! =\! y \equiv \Box
\forall F(xF \equiv yF))$, i.e, abstract objects are identical if and
only if they necessarily encode the same properties. The identity for
ordinary objects on the other hand is classical: $O!x \:\&\: O!y \to
\allowbreak (x\! =\! y \equiv \Box \forall F(Fx \equiv Fy))$, i.e.,
ordinary objects $x$ and $y$ are identical if and only if they
necessarily exemplify the same properties. It is axiomatic that
ordinary objects necessarily fail to encode properties ($O!x \to \Box
\neg \exists FxF)$, and so only abstract objects can be the subject of
true encoding predications.  For example, whereas Pinkerton (a real
American detective) exemplifies \emph{being a detective} and all his other
properties (and doesn't encode any properties), Sherlock Holmes
encodes \emph{being a detective} (and all the other properties
attributed to him in the novels), but doesn't exemplify \emph{being a
  detective}. Holmes, on the other hand, intuitively exemplifies \emph{being
a fictional character} (but doesn't encode this property) and
exemplifies any property necessarily implied by \emph{being abstract}
(e.g., he exemplifies \emph{not having a mass}, \emph{not having a
  shape}, etc.).\footnote{He encodes \emph{having a mass},
  \emph{having a shape}, etc., since these are properties attributed
  to him, at least implicitly, in the story. As an abstract object,
  however, he does \emph{not} exemplify these properties, and so
  exemplifies their negations.}

The key axiom of AOT is the comprehension principle for abstract
objects. It asserts, for every condition on properties (i.e., for
every expressible set of properties), that there exists an abstract
object that encodes exactly the properties that satisfy the condition;
formally:\begin{itemize}
  \item[] $\exists x(A!x \;\&\;\allowbreak \forall F (xF
\equiv\allowbreak \phi))$,\end{itemize}
where $\phi$ is any condition on $F$ in which $x$ doesn't occur
free. Therefore, abstract objects can be modeled as elements of the
power set of properties: every abstract object uniquely corresponds to
a specific set of properties.

Given this basic theory of abstract objects, AOT can define a wide
variety of objects that have been postulated in philosophy or
presupposed in the sciences, including truth-values, Leibnizian
concepts, Platonic forms, possible worlds, natural numbers, and
logically-defined sets.

Another interesting aspect of the theory is its hyperintensionality.
Relation identity is defined in terms of encoding rather than in terms
of exemplification. Two properties $F$ and $G$ are stipulated to be
identical if they are necessarily \emph{encoded} by the same abstract
objects ($F\! =\! G \equiv \Box \forall x(xF \equiv\allowbreak xG)$).
However, the theory does not impose any restrictions on the properties
encoded by a particular abstract object. For example, the fact that an
abstract object encodes the property $[\lambda x\, Fx\; \&\; Gx]$ does
not imply that it also encodes either the property $F$, or $G$ or even
$[\lambda x\, Gx\; \&\; Fx]$ (which, although extensionally equivalent
to $[\lambda x\, Fx\; \&\; Gx]$, is a distinct intensional entity).

Therefore, without additional axioms, pairs of materially equivalent
properties (in the exemplification sense), and even necessarily
equivalent properties, are not forced to be identical. This is a key
aspect of the theory that makes it possible to represent the contents
of human thought much more accurately than classical exemplification
logic would allow.  For instance, the properties \emph{being a
  creature with a heart} and \emph{being a creature with a kidney} may
be regarded as distinct properties despite the fact that they are
extensionally equivalent.  And \emph{being a barber who shaves all and
  only those persons who don't shave themselves} and \emph{being a set
  of all those sets that aren't members of themselves} may be regarded
as distinct properties, although they are necessarily equivalent (both
necessarily fail to be exemplified).

A full description of the theory goes beyond the scope of this
article, but detailed descriptions are available in two books [17],
[18] and various articless by Zalta.  A regularly updated, online
monograph titled \emph{Principia Logico-Metaphysica}~[16] contains the
latest formulation of the theory and serves to compile, in one
location, both new theorems and theorems from many of the published
books and articless.  The mechanization described below follows the
presentation of AOT in a recent version of PLM, but is still being
adapted as research continues and PLM evolves.

The complexity and versatility of AOT, as well as its philosophical
ambitions, make it an ideal candidate to test the universality of the
SSE approach.  However, recent work [13] has posed a challenge
for any embedding of AOT in functional type theory. In the next
section, we briefly discuss this challenge.

\section{AOT in Functional Logic} \label{sec:aotinFL}

 Russell's well-known paradox in naive set theory arises by (a)
 considering the set of all sets that don't contain themselves, and
 (b) noting that this set contains itself if and only if it doesn't.
 A similar construction (`the Clark-Boolos paradox') is possible in
 naive versions of AOT: assume that the term \mbox{$[\lambda x\;
     \exists F (xF\,\&\,\neg Fx)]$} denotes a property, namely, being
 an $x$ that encodes a property that $x$ does not exemplify; call it
 $K$. The comprehension axiom for abstract objects then ensures that
 there is an abstract object that encodes $K$ and no other
 properties. This abstract object then exemplifies $K$ if and only if
 it does not, and so involves one in a paradox.\footnote{Let $a$ be
   the abstract object guaranteed by object comprehension, so that we
   know:\begin{itemize}
  \item[($\vartheta$)] $\forall F(aF \equiv F\! =\! K)$\end{itemize}
Now suppose, for reductio, $Ka$. Then by $\beta$-conversion, there is
a property, say $P$, such that $aP\,\&\,\neg Pa$.  Since $aP$, it
follows by $(\vartheta)$ that $P\! =\! K$. So from $\neg Pa$ it
follows that $\neg Ka$, which contradicts our reductio hypothesis. So
suppose $\neg Ka$.  Then by $\beta$-conversion and predicate logic,
$\forall F(aF \to Fa)$.  Now since $K\! =\! K$, it follows from
$(\vartheta)$ that $aK$. Hence $Ka$. Contradiction.} See [3] for
 details about the paradox; it was first described by
 Clark~[5] and reconstructed independently by Boolos~[2].

AOT undermined the paradox by restricting the matrix of
$\lambda$-expressions to so-called \emph{propositional formulas}, that
is, to formulas without encoding subformulas. This way, the term
\mbox{$[\lambda x\; \exists F(xF\,\&\,\neg Fx)]$} is no longer
well formed and the construction of the paradox fails. Thus, AOT
contained formulas, e.g., $\exists F(xF\,\&\,\neg Fx)$, that may
\emph{not} be placed within a $\lambda$-expression or otherwise
converted to a term.\footnote{In the very latest versions of PLM,
  AOT's free logic has been extended to cover $\lambda$-expressions,
  and so $[\lambda x\; \exists F(xF\,\&\,\neg Fx)]$ is now treated as
  well formed but \emph{non-denoting}.  This latest development is
  briefly discussed in [9].\label{NewAOT}}

Whereas relational type theory allows one to have formulas that cannot
be converted to terms, functional type theory does not; in functional
type theory, it is assumed that every formula can be converted to a
term.  That is crucial to the analysis of the universal
quantifier. The binding operator $\forall x$ in a formula of the form
$\forall x\phi$ is represented, in functional type theory, as a
function that maps the \emph{property} $[\lambda x\:\phi]$ to a truth
value, namely, the function that maps $[\lambda x\:\phi]$ to The True
just in case every object $y$ in the domain is such that $[\lambda x\:
  \phi](y)$ holds.  So in order to represent quantified AOT formulas
that contain encoding subformulas, such as \mbox{$\forall x \exists F
  (xF\,\&\,\neg Fx)$}, their matrices have to be convertible to
terms.\footnote{Note that $\forall x\exists F(xF\,\&\,\neg Fx)$ is a
  well-formed formula of the system, but in fact false. For example,
  it fails when $x$ is ordinary, and when $x$ is the abstract object
  that encodes no properties.  However, the negation of this formula
  is true, and our question is how to \emph{interpret} the embedded
  quantifier in functional type theory.} But, as we've seen, if
\mbox{$[\lambda x\: \exists F (xF \:\&\: \neg Fx)]$} were a term
subject to $\beta$-conversion, AOT would yield a
contradiction.\footnote{Readers familiar with Isabelle/HOL might find
  this notation confusing. In AOT, the symbol $\phi$ is a metavariable
  that ranges over formulas which may contain free occurrences of $x$
  that can be bound by a binding operator.  In Isabelle/HOL, however,
  a formula with a free variable would be represented as a function
  from individuals to truth-values, and the quantified formula would
  be written as $\forall x .\: \phi\: x$. Such a formula is true, if
  $\phi\; x$, i.e., the function application of $\phi$ to $x$ holds
  for all $x$ in the domain.  In this scenario it is true that $\phi =
  (\lambda x .\: \phi\: x)$.  Consequently the primitive, functional
  $\lambda$-expressions of Isabelle/HOL cannot be used to represent
  the $\lambda$-expressions of AOT, since the $\lambda$-expressions of
  Isabelle/HOL cannot simultaneously exclude non-propositional
  formulas while allowing quantified formulas with encoding
  subformulas.}

Thus, it is not trivial to devise a semantical embedding that supports
AOT's distinction between formulas and propositional formulas, but at
the same time preserves a general theory of quantification.  Another
challenge has been to accurately represent the hyperintensionality of
AOT: while relations in AOT are hyperintensional (i.e., necessarily
equivalent relations may be distinct), functions (and relations) in
HOL are fully extensional, and can not be used to represent the
relations of AOT directly.

\section{Embedding AOT in Isabelle/HOL} \label{sec:aotinIsabelle}

The embedding of AOT in Isabelle/HOL overcomes these issues by
constructing a modal, hyperintensional variant of the Aczel-model of
AOT.  Modality is represented by introducing a dependency on primitive
possible worlds in the manner of Kripke semantics of modal
logic. Hyperintensionality is achieved by an additional dependency on
a separate domain of primitive \emph{states}.  Consequently,
propositions are represented as Boolean-valued functions acting on
states and possible worlds.  The model begins with a domain
partitioned into ordinary and \emph{special} urelements (this
corresponds to the domain \textbf{U} in
Figure~\ref{fig:aczel-model}). Whereas the Aczel model represents
properties as sets of urelements, properties in our embedding are
represented as functions mapping urelements to propositions.  The
ordinary objects of AOT are then represented by ordinary urelements,
and the abstract objects of AOT are represented as sets of
properties. These sets are assigned a proxy among the \emph{special}
urelements (and given that there are more sets of properties than
urelements, some abstract objects will be assigned the same
proxy).\footnote{The problem that Aczel solves in the model is this:
  if abstract objects are represented as sets of properties, then how
  are we to understand the fact that in object theory, there is an
  object $x$ and property $F$ such that both $xF$ and $Fx$?  The
  encoding claim, $xF$, is easy: in the model, this is true if $F\in
  x$. However, how can a set of properties \emph{exemplify} a property
  that is an element of it? We cannot model $Fx$ as $x\in F$ without a
  violation of the foundation axiom.  Interestingly, Aczel chose
  \emph{not} to use nonwellfounded sets for his model. Instead he
  mapped abstract objects, modeled as sets of properties, to proxies
  in the domain of special urelements and set the truth conditions for
  $Fx$ to the following disjunctive condition: $x\in F$ if $x$ is
  ordinary and $\|x\| \in F$ (i.e., the proxy of $x$ is an element of
  $F$), if $x$ is abstract.  We have extended Aczel's model with
  modality and hyperintensionality.}

\tikzset{font=\fontsize{8pt}{10pt}\selectfont}
\begin{figure}
\centering
\begin{tikzpicture}

% Domains
 \node at (-2.4,-1) {Domain \textbf{D} = $\mathbf{A} \cup \mathbf{C}$};

% \node at (-.8,-1.5) {Define for $\mathitbf{x}\in \textbf{D}$,
%   $|\mathitbf{x}| =
%   \left\{\begin{array}{ll} \hspace*{-.05in}\mathitbf{x},
%   \textrm{when}\ \mathitbf{x}\in
%   \mathbf{C}\\ \hspace*{-.05in}\|\mathitbf{x}\|,
%   \textrm{when}\ \mathitbf{x}\in \mathbf{A} \end{array} \right.$};

% U
 \draw (0,0) ellipse (1 and .6);
 \draw (0,.6) -- (0,-.6);
 \node at (-2.5,0) {\textbf{U} = Urelements =};
 \node at (2.5,.7) {Define a mapping:};
 \node at (2.5,.4) {$\|a\| : \textbf{A} \to \textbf{S}$};
 \node at ($(-.45,.8)+(-90:1 and .6) + (0,-.2)$) {\textbf{C}};
 \node at ($(.45,.8)+(-90:1 and .6) + (0,-.2)$) {\textbf{S}};
 \node (S) at ($(.45,.9)+(-90:1 and .6) + (0,-.2)$) {};

% P
 \draw (0,2) ellipse (1.7 and 1);
 \node at (0,2) {\textbf{P} = Properties = $\wp (\mathbf{U})$};

% A
 \draw (0,4.8) ellipse (2.2 and 1.3);
 \node at (0,4.8) {\textbf{A} = Abstract Objects = $\wp (\mathbf{P})$};
 \node (A) at (1.7,4.1) {};

% Arrows
 \draw [>->] (A) to[out=-45, in=40] (S);

\end{tikzpicture}
\caption{Extensional, non-modal Aczel model of
  AOT.}\label{fig:aczel-model}
\end{figure}
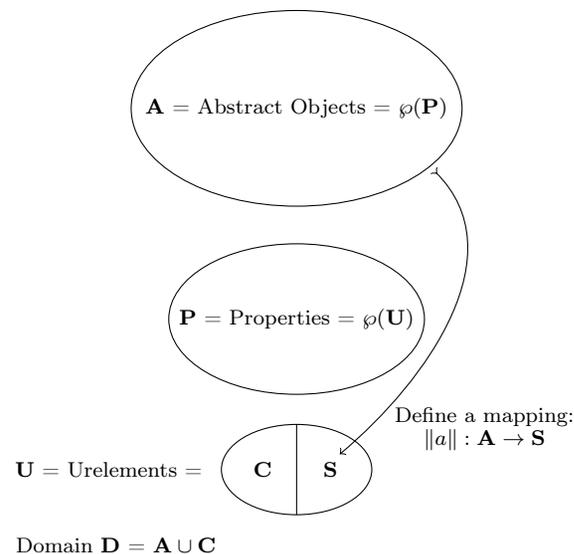

\begin{figure}[ht]
\includegraphics[width=.7\textwidth]{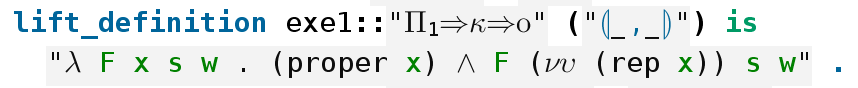}
\includegraphics[width=1.\textwidth]{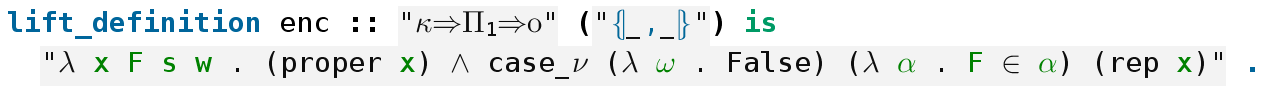}

\caption{Definition of 1-place exemplification and encoding in
  Isabelle/HOL. }\label{fig:exe_and_enc}
\end{figure}

From this description, it becomes clear that if $x$ is an ordinary
object, then the truth conditions of an exemplification formula $Px$
are captured by the proposition that is the result of applying the
function representing the property $P$ to the ordinary urelement
representing $x$. If $x$ is an abstract object, then the truth
conditions of an exemplification formula $Px$ are captured by the
proposition that is the result of applying the function representing
the property $P$ to the \emph{special} urelement that serves as the
proxy of $x$.  An encoding formula $xP$, by contrast, is true just in
case $x$ is an abstract object and the property $P$ is an element of
the set of properties representing $x$.  This latter feature of the
model validates the comprehension axiom for abstract objects: for
every set of properties, there exists a unique abstract object that
encodes exactly those properties. Figure \ref{fig:exe_and_enc} shows
the representation of 1-place exemplification and encoding in
Isabelle/HOL.

In the definition of 1-place exemplification, which starts on the
first line, \textsf{exe1} is defined as a function of type
${\Pi_1}\!\Rightarrow\!\kappa\!\Rightarrow\! o$ that maps 1-place
relation terms (of type $\Pi_1$) and individual terms (of type
$\kappa$) to propositions (type $o$). This function then becomes
represented as a 4-argument function that maps properties,
individuals, states, and worlds to a Boolean (this function, in turn,
is defined by means of \textsf{proper}, \textsf{rep} and the
$\nu\upsilon$ mappings).\footnote{Here \textsf{proper} $x$ is true, if
  $x$ denotes an individual ($x$ can also be a non-denoting definite
  description), \textsf{rep} $x$ is the individual denoted by $x$
  (given that $x$ denotes), and $\nu\upsilon$ is the mapping from
  individuals to urelements. As a result, the exemplification function
  maps a property term and an individual term to a proposition that is
  true in a given intensional state and possible world if and only if
  the individual term denotes and the property denoted by the property
  term maps the triple consisting of the urelement corresponding to
  the denoted individual, the given intensional state, and the given
  possible world, to The True. For a full description of all types,
  symbols and concepts involved, refer to~[8].} In the
definition of encoding, \textsf{enc} is defined as a function of type
$\kappa\!\Rightarrow\!{\Pi_1}\!\Rightarrow\! o$ that maps individual
terms and 1-place relation terms to propositions. This function then
becomes represented as a 4-argument function that maps individuals,
properties, states, and worlds to a Boolean (this function, in turn,
is defined by means of \textsf{proper}, \textsf{rep} and a case
distinction on types \textsf{case}\_$\nu$).\footnote{The second
  conjunct in the definition evaluates to \textsf{False}, if
  \textsf{rep} $x$ is an ordinary object; if \textsf{rep} $x$ is an
  abstract object, the conjunct is true if and only if the property
  $F$ is contained in the abstract object.}

Since well-formed $\lambda$-expressions in AOT were required to have a
propositional matrix, such expressions correspond to functions on
urelements.  Given that encoding subformulas were excluded from these
expressions in AOT, the only formulas that can occur in the matrix of
a $\lambda$-expression are those built up from exemplification
formulas. The truth conditions of these formulas are determined solely
by the properties and relations of the urelements in the
model.\footnote{It turns out that in the October 28, 2016 version of
  \emph{PLM}, there was an exception to this rule that led to the
  reintroduction of the Clark-Boolos paradox. We'll discuss this in a
  subsequent section.}

\begin{figure}[ht]
\hspace*{.25in}\includegraphics[width=0.9\textwidth]{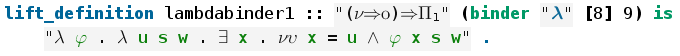}
\caption{Definition of AOT's $\lambda$-expressions in Isabelle/HOL.}\label{fig:lambda}
\end{figure}

Consequently, the $\lambda$-expressions of AOT are not represented
using the unrestricted primitive $\lambda$-expressions of HOL, but
have a more complex semantic representation which is captured by the
definition of a new class of $\lambda$-expressions in Isabelle/HOL
that will represent AOT $\lambda$-expressions; the definition for the
1-place case is given in Figure \ref{fig:lambda}.

It may help to explain the definition of \textsf{lambdabinder1} in
Figure \ref{fig:lambda}. \textsf{lambdabinder1} is defined as a
function of type $(\nu \Rightarrow o) \Rightarrow \Pi_1$, which maps
functions-from-individuals-to-propositions to 1-place relation
terms. The function $\varphi$ is mapped to a 1-place relation term,
which is in turn represented as a Boolean-valued ternary function on
urelements $u$, states $s$, and worlds $w$. This function evaluates to
true if there exists an individual $x$ such that both $x$ is mapped to
the urelement $u$ under the mapping $\nu\upsilon$ and the function
$\varphi$ evaluates, for $x$, to a proposition true in $s$ and $w$.

Thus, non-well-formed $\lambda$-expressions of AOT, which can't be
syntactically excluded from the SSE representation, are given a
nonstandard semantics and this avoids, modulo the discussion below,
the Clark-Boolos paradox.  As a result, $\beta$-conversion for the
\emph{defined} $\lambda$-expressions holds in general for terms that
were syntactically well-formed in AOT, whereas for terms that were not
syntactically well-formed in AOT (but which are still part of the
SSE), $\beta$-conversion is not derivable.

The model structure we've just described can represent all the terms
of the target logic and can preserve hyperintensionality. Moreover,
the axioms and inference rules of AOT become derivable. Thus, the
embedding makes it possible to introduce additional layers of
abstraction.  Given the model structure, the first layer of
abstraction is the representation of the formal semantics of
PLM. This, in turn, becomes the basis of the second layer of
abstraction, namely, the derived axioms and the fundamental inference
rules of PLM.  The second abstraction layer consists solely of the
axioms and rules of PLM itself and makes it possible to reason
directly in the target logic but independently of the underlying model
structure.  Thus, the second layer of abstraction avoids the
derivation of artifactual theorems; the fact that the model structure
validates formulas that aren't theorems of AOT is of no further
consequence.  And, just as importantly, the model guarantees that the
system of AOT is sound.

\begin{figure}[h]
\centering
\includegraphics[width=\textwidth]{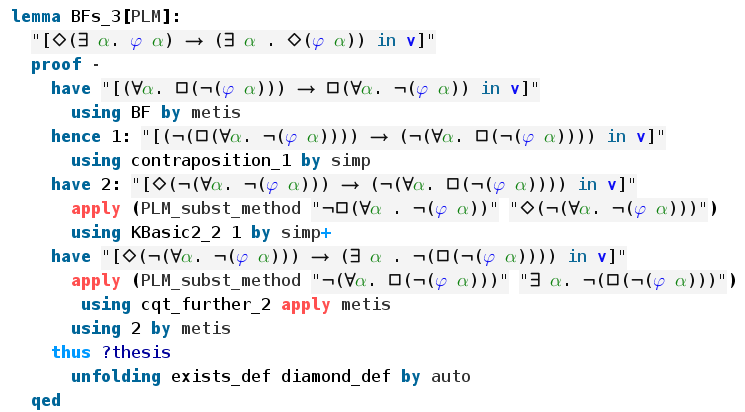}
\caption{Reasoning in the abstract layer in Isabelle/HOL. Only
  theorems and rules of AOT are used in
  derivations.}\label{fig:barcan}
\end{figure}

\begin{figure}[h]
\includegraphics[width=0.55\textwidth]{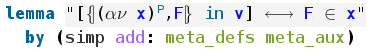}
\caption{Artifactual theorems are provable only by expanding the
  metalogical definitions.}\label{fig:artifactual}
\end{figure}

These results are illustrated in the following figures.  In Figure
\ref{fig:axioms}, we show the derivation of some axioms; in Figure
\ref{fig:barcan}, we show the derivation of the $\Diamond$ version of
the Barcan formula; and in Figure \ref{fig:artifactual}, we show that
the derivation of the artifactual theorem $xF \leftrightarrow F\in x$
requires one to unfold the semantic definitions (this is revealed by
the second line).

\begin{figure}[h]
\includegraphics[width=0.7\textwidth]{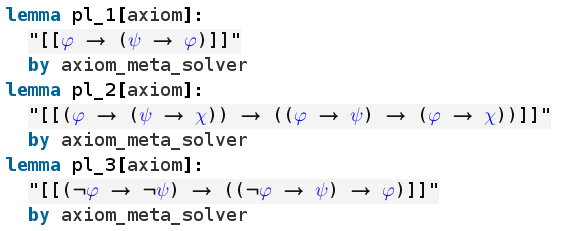}
\caption{Some of the axioms of AOT, derived automatically in
  Isabelle/HOL.}\label{fig:axioms}
\end{figure}

Furthermore, there are other advantages to our methodology.  For one
thing, it is straightforward to convert statements derived within
Isabelle/HOL into traditional pen and paper proofs for AOT.  Thus, our
approach facilitates experimental studies within the computational
implementation and informs discussions about them. Moreover, the
approach is suitable for conducting a deeper analysis of AOT and its
model structure. The analysis led to the discovery of how a
previously-known paradox could easily resurface if care isn't taken in
the formulation of PLM. This paradox will be sketched in the next
section.

\section{Reintroduction of a Paradox} \label{sec:paradox}

As explained in the previous section, our goal was to ensure that all
of the $\lambda$-expressions of the embedding that conformed to AOT's
syntactic restrictions have a standard semantics. We wanted
$\beta$-conversion to govern all $\lambda$-expressions with a
propositional matrix.

However, as we analyzed our work, it became apparent that in the
working version of PLM that served as the basis of our investigations,
certain $\lambda$-expressions involving definite descriptions did not
exhibit the desired behavior in the embedding. These were definite
descriptions containing a free occurrence of a variable that becomes
bound by a $\lambda$-binder.  We could not verify that
$\beta$-conversion was derivable for those expressions. Using the
infrastructure provided by Isabelle/HOL, it became possible to show
that $\beta$-conversion for these terms does not hold generally in an
Aczel model and this suggested that there might be some problem with
these expressions.

Consequently, we focused our attention on the assumption that
$\beta$-conversion holds for such terms in Isabelle/HOL. This
assumption turned out to be inconsistent; the layered structure of the
embedding made it possible to construct a proof of the inconsistency
using object-level reasoning at the highest level of abstraction. This
way, a human-friendly proof of the paradox was reconstructed and
quickly confirmed. It became apparent that the logic of
$\lambda$-expressions and definite descriptions combined to circumvent
the restriction that encoding subformulas not be allowed in
$\lambda$-expressions. Indeed, the paradox turned out to be one that
was previously known (namely, the Clark-Boolos paradox mentioned
earlier), but which had re-emerged through the back door (see the
discussion below). This new route to a previously-known paradox
constituted a new paradox.  The new paradox is due in part to the
precise definition of \emph{subformula}. The matrix of a
$\lambda$-expression in AOT was allowed to contain encoding formulas
as long as they were \emph{nested within a definite description}.
Encoding formulas so nested were not considered subformulas of the
matrix and so such matrices were still considered propositional
formulas. Therefore, the term \mbox{$[\lambda x\; G\iota y\psi]$} was
considered well formed even if $\psi$ contained encoding
subformulas. By choosing $G$ to be a property that is universally true
(e.g. \mbox{$[\lambda z\; \forall p (p \rightarrow p)]$}) and letting
$\psi$ be \mbox{$y = x \:\&\: \exists F(xF \:\&\: \neg Fx)$}, one
could construct a property that is extensionally equivalent to the
property $K$ described above in Section~\ref{sec:aotinFL}. This was
sufficient to reconstruct the Clark-Boolos paradox.

More specifically, to see how the Clark-Boolos paradox was
reintroduced, suppose that the following $\lambda$-expression denotes
a property, for any choice of $G$: \begin{enumerate}
  \item[] $[\lambda x\: G\imath y(y\! =\! x \:\&\: \exists F(xF \:\&\:
    \neg Fx))]$\end{enumerate}
Then if $G$ is a universal property such that $\forall xGx$, it can be
shown that
\begin{enumerate} 
\item[] \begin{tabular}{lll} $G\imath y(y\! =\! x \:\&\: \exists
  F(xF \:\&\: \neg Fx))$ & $\equiv$ & $\exists !y(y\! =\! x \:\&\:
  \exists F(xF \:\&\: \neg Fx))$ \\ & $\equiv$ & $\exists F(xF \:\&\:
   \neg Fx)$\end{tabular} \end{enumerate}
We leave the proof as an exercise.  So the matrix $G\imath y(y\! =\! x
\:\&\: \allowbreak \exists F(xF \:\&\: \allowbreak \neg Fx))$ is
equivalent to $\exists F(xF \:\&\:\allowbreak \neg Fx)$, when $G$ is a
universal property.  Although the $\lambda$-expression built from the
latter matrix was banished from AOT, a $\lambda$-expression built from
the former matrix would just as easily lead to the Clark-Boolos
paradox.

The discovery of the reemergence of the Clark-Boolos paradox has led
to some new developments in AOT.  There has been a modest revision of
the axioms of AOT which avoids the paradox (without sacrificing any
important theorems).  Indeed, as mentioned earlier in
Footnote~\ref{NewAOT}, AOT now allows encoding subformulas in
$\lambda$-expressions, but its free logic ensures that only the
\emph{safe} ones denote properties.  These developments also led to a
new definition of \emph{logical existence} for terms, i.e., the fact
that a term has a denotation and is thus significant.  Typically, in
free logic, one defines the logical existence of a term (represented
as $\tau\!\!\downarrow$) as $\exists \beta (\beta\! =\! \tau)$. Object
theory had followed this pattern for individual terms. Now, logical
existence is (a) defined using exemplification predication instead of
using identity for individual terms and (b) also defined for relation
terms, this time using encoding predication. These new developments of
AOT and its embedding are briefly discussed in~[9].

\section{Final Considerations} \label{sec:final-considerations}

The complexity of the target system and the multiple abstraction
layers presents a challenge for the development and use of automated
reasoning tools. Normally, one would automate proofs of any embedded
theory by using Isabelle/HOL's inbuilt reasoning tools (e.g.,
Sledgehammer and Nitpick) to unfold the semantical definitions used to
represent the theory in HOL.  But, in the case of PLM, a better option
is to directly automate the proof theory as an abstraction layer,
i.e., without unfolding the semantical definitions.  We had two
reasons for adopting this latter option: it easily avoids the problem
of generating artifactual theorems and it allows for the interactive
construction of complex, but human-friendly, proofs for PLM.  To
simplify the implementation of this option, we used the \emph{Eisbach
  package} of Isabelle to define proof methods for the system PLM,
including a resolution prover that can automatically derive the
classical propositional tautologies directly in AOT.

One interesting problem that has not yet been resolved is the one
identified in Oppenheimer \&\ Zalta~[13]. As noted in
Section~\ref{sec:aotinFL}, AOT has formulas that can't be converted to
denoting terms and this makes it difficult to give a general
representation of AOT in functional type theory.  Oppenheimer
\&\ Zalta concluded from this that relational type theory is more
fundamental than functional type theory.  But though the SSE embedding
of AOT in Isabelle/HOL doesn't challenge this conclusion directly, it
does show that the functional setting of HOL can offer a reasonably
accurate representation of the reasoning that can be done in AOT.
This approach addresses, at least in part, Oppenheimer \&\ Zalta's
claim, though we haven't yet addressed whether functional type theory,
in the absence of abstraction layers, can generally represent systems
of relational type theory in which not every formula can be converted
to a denoting term.

We've discovered that the key to the development of a sound
axiomatization of the complex relation terms of AOT is to be found in
the study of, and solution to, the representation of
$\lambda$-expressions. With a paradox-free emendation of AOT, current
research is directed to giving an extended analysis of the
faithfulness of the embedding approach we used; this would shed
further light on the debate about relational and functional type
theory.  This study should be complemented by an analysis of the
reverse direction, i.e., an embedding of the fundamental logic of HOL
in the (relational) type-theoretic version of AOT.  Both studies
should then be carefully assessed.

In conclusion, the semantical embedding approach has been fruitfully
employed to encode the logic of AOT in Isabelle/HOL. By devising and
utilizing a multi-layered approach (which at the most abstract level
directly mechanizes the proof-theoretic system of AOT), the issues
arising for an embedding in classical higher-order logic are not too
difficult to overcome.  A highly complex target system based on a
fundamentally different tradition of logical reasoning (relational
instead of functional logic) has been represented and analyzed using
the approach of shallow semantical embeddings.  The power of this
approach has been demonstrated by the discovery of a previously
unnoticed paradox that was latent in AOT.  Furthermore, the work
contributes to the philosophical debate about the tension between
functional type theory and relational type theory and their
inter-representability, and it clearly demonstrates the merits of
\emph{shallow semantical embeddings} as a means towards universal
logical reasoning.

\section*{References}

\begin{description}

% \bibfitem{J41}
\item[{[}1{]}] Christoph Benzm{\"u}ller, \emph{Universal
      (meta-)logical reasoning: Recent successes}, \textbf{Science of
      Computer Programming}, Volume 172 (2019), 48--62.

% \bibfitem{cffa}
\item[{[}2{]}] George Boolos, {\em The Consistency of Frege's
    Foundations of Arithmetic}, in J.~Thomson (ed.), \textbf{On Being
    and Saying}, Cambridge, MA: MIT Press, 1987, pp.~3--20.

% \bibfitem{wpsf}
\item[{[}3{]}] Ot\'avio Bueno, Christopher Menzel, and
  Edward~N. Zalta, {\it Worlds and Propositions Set Free},
  \textbf{Erkenntnis}, 79 (2014): 797--820.

% \bibfitem{Church40}
\item[{[}4{]}] Alonzo Church, {\em A Formulation of the Simple Theory
    of Types}, \textbf{Journal of Symbolic Logic}, 5 (1940): 56--68.

% \bibfitem{neot}
\item[{[}5{]}] Romane Clark,  {\em Not Every Object of Thought has
  Being: A Paradox in Naive Predication Theory}, \textbf{No\^us},
  12 (1978): 181--188. 

% \bibfitem{Begriffsschrift}
\item[{[}6{]}] Gottlob Frege, \textbf{Begriffsschrift, eine der
  arithmetischen nachgebildete Formelsprache des reinen Denkens},
  Halle: Verlag von Louis Nebert, 1879.

% \bibfitem{sep-mally}
\item[{[}7{]}] Alexander Hieke and Gerhard Zecha, {\it Ernst Mally},
  \textbf{The Stanford Encyclopedia of Philosophy} (Winter 2018
  Edition), Edward~N. Zalta (ed.), URL =\\
  https://plato.stanford.edu/archives/win2018/entries/mally/

% \bibfitem{Thesis}
\item[{[}8{]}] Daniel Kirchner,  {\it Representation and Partial
  Automation of the Principia Logico-Metaphysica in Isabelle/HOL},
  \textbf{Archive of Formal\\ Proofs}, 2017, URL = 
  http://isa-afp.org/entries/PLM.html

% \bibfitem{KBZ1}
\item[{[}9{]}] Daniel Kirchner, Christoph Benzm\"uller, and
  Edward~N. Zalta, {\it Computer science and metaphysics: A
    cross-fertilization}, \textbf{Open Philosophy} (Special Issue --
    Computer Modeling in Philosophy), Patrick Grim (ed.), 
    forthcoming, doi: 10.1515/opphil-2019-0015, preprint available at
   https://arxiv.org/abs/1905.00787

% \bibfitem{sep-principia-mathematica}
\item[{[}10{]}] Bernard Linsky and Andrew~David Irvine, 
  {\it Principia Mathematica},
  \textbf{The Stanford Encyclopedia of Philosophy} (Summer 2019 Edition),
  Edward~N. Zalta (ed.), URL =  
  https://plato.stanford.edu/\\archives/sum2019/entries/principia-mathematica/

% \bibfitem{Muskens2007}
\item[{[}11{]}] Reinhard Muskens, {\em Intensional models for the
  theory of types}, \textbf{Journal of Symbolic Logic}, 72 (2007):
  98--118.

% \bibfitem{Isabelle}
\item[{[}12{]}] Tobias Nipkow, Lawrence~C. Paulson, and Markus Wenzel,
  \textbf{Isabelle/HOL --- a proof assistant for higher-order logic}
  (Lecture Notes in Computer Science: Volume 2283), Berlin, Heidelberg:
  Springer, 2002.

% \bibfitem{rtt}
\item[{[}13{]}] Paul~E. Oppenheimer and Edward~N. Zalta,
  \emph{Relations Versus Functions at the Foundations of Logic: Type-Theoretic
  Considerations}, \textbf{Journal of Logic and Computation},
  21 (2011): 351--374.

% \bibfitem{Whitehead-Russell-1913}
\item[{[}14{]}] Alfred North Whitehead and Bertrand Russell,
 \textbf{Principia Mathematica}, 1st edition, Cambridge
  University Press, Cambridge, 1910-1913.

% \bibfitem{wiedijk:_bruij}
\item[{[}15{]}] Freek Wiedijk, {\em The ``de Bruijn factor''}, website,
  2012, URL =\\ http://www.cs.ru.nl/F.Wiedijk/factor/
 (accessed July 23, 2019).

% \bibfitem{PM}
\item[{[}16{]}] Edward~N. Zalta, {\em Principia Logico-Metaphysica},
  online monograph, 2018, URL =
  http://mally.stanford.edu/principia.pdf (accessed: April 11, 2018).

% \bibritem{zalta1983abstract}
\item[{[}17{]}] ---------, \textbf{Abstract Objects: An Introduction to
  Axiomatic Metaphysics}, Dordrecht: D.~Reidel, 1983, available online 
   URL = http://mally.stanford.edu/abstract-objects.pdf

% \bibritem{zalta1988intensional}
\item[{[}18{]}] ---------, \textbf{Intensional Logic and the Metaphysics
  of Intentionality}, Cambridge, MA: MIT Press, 1988, available online
  URL = http://mally.stanford.edu/intensional-logic.pdf

\end{description}

\end{document}